\documentclass[reprint,showpacs,preprintnumbers,amsmath,amssymb,aps,prl]{revtex4-1}
\usepackage{graphicx}% Include figure files
\usepackage{color}
\usepackage{dcolumn}% Align table columns on decimal point
\usepackage{bm}% bold math
\usepackage{multirow}
\usepackage{bbding}

\allowdisplaybreaks[1]

  \newcommand{\average}[1]{\ensuremath{\langle#1\rangle} }

\begin{document}

\title{Helicity current generation by distorted Rashba coupling}
\author{Takumi Funato}
\affiliation{%
Kavli Institute for Theoretical Sciences, University of Chinese Academy of Sciences, Beijing, 100190, China.
}
\affiliation{Center for Spintronics Research Network, Keio University, Yokohama 223-8522, Japan}
\author{Mamoru Matsuo}
\affiliation{%
Kavli Institute for Theoretical Sciences, University of Chinese Academy of Sciences, Beijing, 100190, China.
}%
\affiliation{%
CAS Center for Excellence in Topological Quantum Computation, University of Chinese Academy of Sciences, Beijing 100190, China
}%
%\email{}
\affiliation{RIKEN Center for Emergent Matter Science (CEMS), Wako, Saitama 351-0198, Japan}
\affiliation{
Advanced Science Research Center, Japan Atomic Energy Agency, Tokai, 319-1195, Japan
}

\date{\today}

\begin{abstract}
We theoretically study spin transport in two-or three-dimensional Rashba systems dynamically distorted by surface acoustic waves. The spin currents in the linear response to lattice distortion dynamics are calculated on the basis of a microscopic theory combined with local coordinate transformations. As a result, we find a mechanism of direct spin-current generation from lattice distortion not associated with a charge current or spin accumulation.
Moreover, the in-plane helicity currents are generated by shear surface acoustic waves via the present mechanism. The generated helicity currents are not parallel to the vorticity of the lattice, and cannot be created with the conventional methods. Thus, our findings offer an alternative functionality of the conventional Rashba systems in the field of spintronics. 
\end{abstract}

\pacs{Valid PACS appear here}
\maketitle

%intro
In the field of spintronics, spin current creation and manipulation is a central research topic.
The interconversion between spin and charge via spin-orbit interaction (SOI) plays a key role specifically the direct and inverse spin Hall effects (SHE)\cite{d'yakonov1971,hirsch1999,zhang2000,murakami2003,sinova2004,engel2005,saitoh2006,sinova2015}, Edelstein effect\cite{edelstein1990,li2014,fan2014,karube2016,manchon2015}, and spin galvanic effect\cite{ganichev2002} among others.

Recently, spin-current generation due to the conversion of macroscopic angular momentum associated with mechanical motion (e.g, rigid rotation, vibration, and vortex) into spin angular momentum has been considered.
In particular, spin current generation via spin-vorticity coupling (SVC) has received broad attention because it does not require SOI.  
SVC is a coupling of electron spin and effective magnetic fields arising in the rotating (non-inertial) frame which is locally fixed on the moving materials\cite{matsuo2011,matsuo2013}.
Takahashi {\it et al.} measured electric voltage in the shear flow of liquid metal, and proposed that the spin current was generated from the vorticity motion of the liquid metal via SVC\cite{svc_expt1,svc_expt2,svc_expt3,svc_theo1}.
Kobayashi {\it et al.} observed spin-torque spin-wave resonance due to an AC spin current driven by the lattice vorticity motion associated with surface acoustic waves (SAWs) via SVC\cite{kobayashi2017,tateno2020}.

The effect of SOI on mechanical spin-current generation has also been studied.
Kawada {\it et al.} applied SAWs to a heavy metal/ferromagnetic metal bilayer under an in-plane magnetic field and investigated the angular dependence of DC voltage on the in-plane magnetic field\cite{kawada2021}.
They found characteristic magnetic dependence, which was not seen when using weak SOI metal instead of heavy metal
, indicating mechanical spin-current generation via intrinsic SOI.
Moreover, a previous work theoretically studied a spin current driven by lattice distortion dynamics via extrinsic SOI,
and presented two mechanisms: one via the SHE from the charge current and the other via the diffusion of the spin accumulation\cite{funato2018}.
It has also been suggested that spin current generation via extrinsic SOI is comparable with that via SVC in strong SOI systems. 
Nonetheless, the effects of intrinsic SOI remain an open problem.
Direct spin-current generation by mechanical means, e.g, via modulation of intrinsic SOI, is expected.

\begin{figure}[bp]
\centering
\includegraphics[width=75mm]{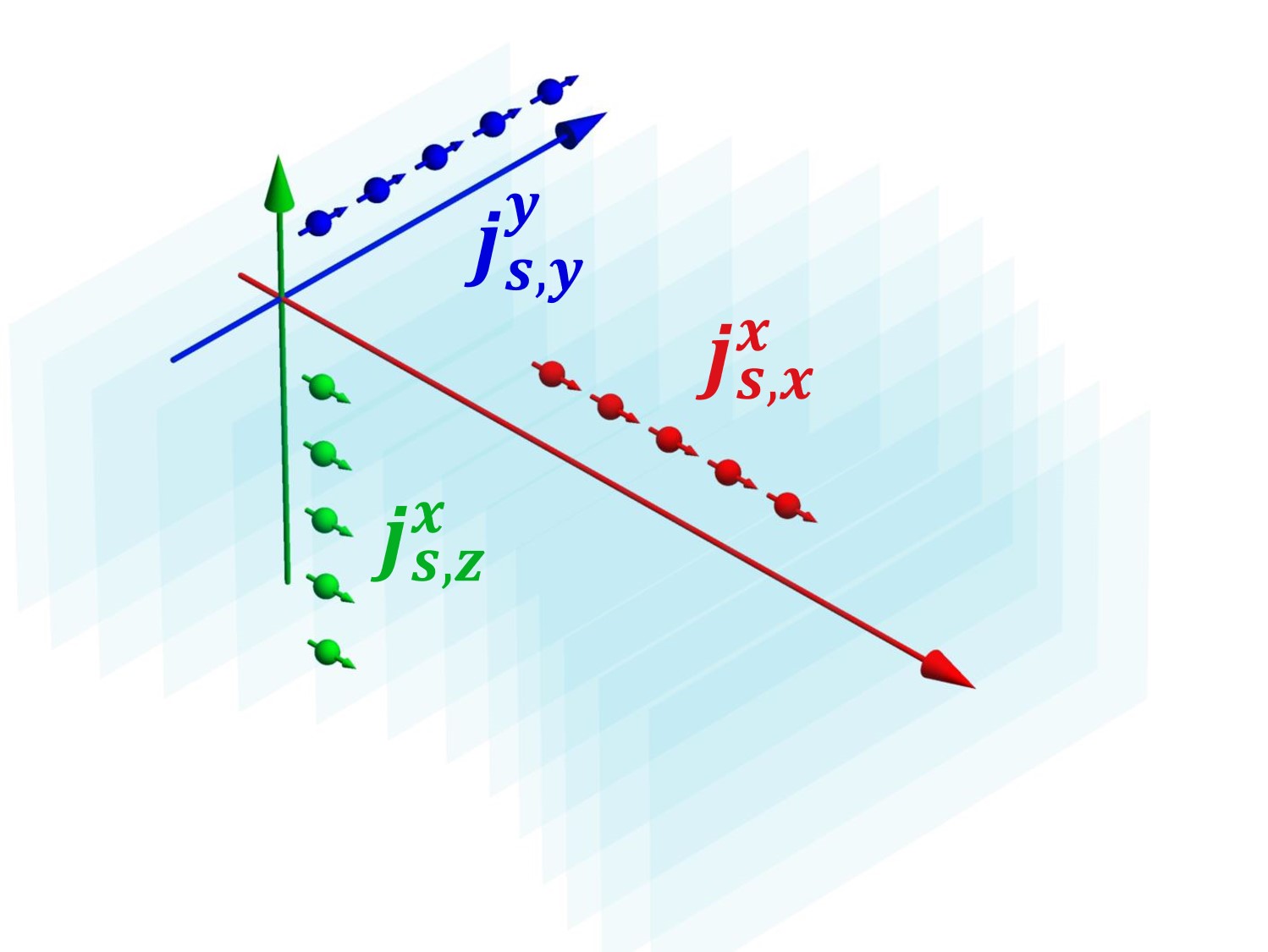}
\caption{
Helicity currents $j^x_{\text s,x}$, $j^y_{\text s, y}$ are generated by the SSAW in Rashba systems. 
In three-dimensional Rashba systems, the longitudinal spin current $j^x_{\text s, z}$ is also induced.}
\label{wave}
\end{figure}

The purpose of this research is to explore the mechanisms of mechanical spin-current generation involved in intrinsic SOI.
As a first step, we focused on Rashba SOI\cite{rashba1,rashba2} originating from spatial inversion symmetry breaking.
We considered a Rashba electron system, in which lattice distortion dynamics are induced.
Treating the effect of lattice distortion with local coordinate transformation, we calculated the non-equilibrium spin current in the linear response to lattice displacement.
The results suggested that the spin current was generated directly from the lattice displacement instead of via the charge current or spin accumulation. 
A helicity current flows in the in-plane direction by shear SAW (SSAW), which could not be created via conventional methods, such as SVC and SHE.
The spin current generated by the present mechanism is large enough to be observable and expected to be experimentally observed using spin wave resonance.

%model
We consider two- and three-dimensional electrons subjected to impurity potential in the presence of the lattice periodic potential and its SOI:
\begin{align}
H = \frac{p^2}{2m} + V_{\text p}(\bm r') + V_{\text i} (\bm r') + \lambda _{\text{so}}\bm \sigma \cdot [\nabla ^{\prime} V_{\text p} (\bm r') \times \bm p], 
\end{align}
where $V_{\text p}(\bm r')$ is the periodic potential, $V_{\text i}(\bm r')$ is the impurity potential, $\lambda _{\text{so}}$ is the SOI strength, and $\bm \sigma = (\sigma ^x,\sigma ^y,\sigma ^z)$ are the Pauli matrices.
The point like impurity potentials, $V_{\text i}(\bm r')=u_{\text i}\sum_j \delta (\bm r'-\bm R_j)$, are assumed where $u_{\text i}$ is the strength of the impurity potential  and $\bm R_j$ is the position of the $j$-th impurity.
When lattice distortion dynamics were applied, e.g, because of sound waves and external forces, the Hamiltonian is deformed as 
\begin{gather}
H_{\text T} = \frac{p^2}{2m} + V_{\text p}(\bm r'-\delta \bm R) + V_{\text i} (\bm r'-\delta \bm R) \nonumber \\
+ \lambda _{\text{so}}\bm \sigma \cdot [\nabla ^{\prime} V_{\text p} (\bm r'-\delta \bm R) \times \bm p], 
\end{gather}
where $\delta \bm R(\bm r', t)$ is the displacement vector of the lattice.

To treat the lattice distortion, we perform a local coordinate transformation, $\bm r=\bm r'-\delta \bm R(\bm r', t)$, from the laboratory frame (with $\bm r'$) to a "material" frame (with $\bm r$) locally fixed to the deformed lattice.
We assume the free-electron systems with $z-$direction spatial symmetry breaking.
Up to the first order in $\delta \bm R$, the total Hamiltonian is given by 
\begin{align}
H_{\text T} = H + H'(t).
\end{align}
The first and second terms are a time-independent Hamiltonian and an external perturbation Hamiltonian, respectively, due to the lattice distortion dynamics.
Time-independent term $H=H_0+H_{\text{i}}$ consists of an unperturbed (kinetic energy and Rashba  SOI) term $H_0=\sum_{\bm k}c^{\dagger}_{\bm k}[\frac{k^2}{2m} + \alpha _{\text R}\hat z\cdot (\bm k\times \bm \sigma )]c_{\bm k}$ and an impurity potential term $H_{\text i}=\sum_{\bm k,\bm k'}V_{\bm k'-\bm k}c^{\dagger}_{\bm k'}c_{\bm k}$, 
where $c_{\bm k}^{\dagger}(c_{\bm k})$ is the creation (annihilation) operator of the electrons, $\alpha _{\text R} = \lambda _{\text{so}}\int d^3r \partial _z V_{\text p}$ is the strength of the Rashba SOI,
and $V_{\bm k'-\bm k}$ is the Fourier component of the impurity potential.

The external perturbed Hamiltonian due to the lattice distortion dynamics consists of two parts, $H'(t)=H'_{\text K}(t)+H'_{\text R}(t)$ (the detailed derivation can be found in the Supplementary Material):
\begin{gather}
H'_{\text K}(t) = -\sum_{\bm k} c^{\dagger}_{\bm k+} W^{\text K}_j(\bm k) c_{\bm k-} u^j_{\bm q,\omega},
\\
H'_{\text R}(t) = -\sum_{\bm k} c^{\dagger}_{\bm k_+} W_j ^{\text R}(\bm k) c_{\bm k_-} u^j_{\bm q, \omega},
\end{gather}
where $\bm k_{\pm}=\bm k\pm \frac{\bm q}{2}$ and $\bm u_{\bm q,\omega}=-i\omega \delta \bm R_{\bm q,\omega}$ are the Fourier components of the lattice velocity field.
Here, $W^{\text K}_j$ and $W^{\text R}_j$ are given by
\begin{gather}
W^{\text K}_j(\bm k) = \left( 1-\frac{\bm v\cdot \bm q}{\omega} \right) k_j,
\\
W^{\text R}_j(\bm k) = \frac{\alpha _{\text R}}{\omega} \left[
\bm k\cdot (\bm \sigma \times \bm q) \delta _{jz} -\hat z \cdot (\bm \sigma \times \bm q) k_j 
\right].
\end{gather}
Throughout this study, we assume that the wavenumber $q$ and the frequency $\omega$ are much smaller than inverse of mean-free path $l$ and relaxation time $\tau$ of electrons, $q\ll l^{-1}$ and $\omega \ll \tau ^{-1}$, respectively.

Assuming uniform random distributed impurities, the impurity averaged equilibrium retarded/advanced Green functions are given by 
\begin{align}
G^{\text{R/A}}_{\bm k}(\epsilon ) = \frac{1}{2}\sum_{s=\pm} \frac{1+s\hat \Gamma _{\bm k}}{\mu +\epsilon -E_{(s)}\pm i\gamma},
\end{align}
where $E_{(s)}=\frac{k^2}{2m}+s\alpha _{\text R}k_{\text t}$ with $k_{\text t}=(k_x^2+k_y^2)^{\frac{1}{2}}$ is the eigenenergy, 
$\mu$ is the chemical potential, and $\hat \Gamma _{\bm k}=\epsilon _{\beta l} \sigma ^{\beta} \frac{k_l}{k_{\text t}}$ with $\epsilon _{\beta l}=\epsilon _{z\beta l}$ being the Levi-Civita symbol. 
Here, $\gamma = \frac{\pi}{2}n_{\text i} u_{\text i}^2\nu (\mu )$ is the damping rate calculated in the Born approximation, where $n_{\text i}$ is the impurity concentration and $\nu (\mu )=\nu _0 a_1$ is the Fermi-level density of states, 
with $\nu _0=\frac{mk_{\text F}}{2\pi ^2}$(for 3D) and $\frac{m}{2\pi}$ (for 2D) being the density of states in free-electron systems without Rashba SOI, $k_{\text F}=\sqrt{2m\mu}$ being Fermi wavenumber, and
$a_1$ being a dimensionless parameter defined after Eq.~(\ref{result}).
Here, we use the ensemble average for impurity positions $\average{V_{\bm k}V_{\bm k'}}_{\text{av}}=n_{\text i}u_{\text i}^2\delta _{\bm k,\bm k'}$.

The Spin current operator is given by
\begin{align}
\hat j^{\alpha}_{\text s, i}(\bm q) = \sum_{\bm k} c^{\dagger}_{\bm k_-} j^{\alpha}_{\text s, i} c_{\bm k_+},
\end{align}
where $\alpha$($=x,y,z$) specifies the spin direction, $i$($=x,y,z$) specifies the flow direction, and $j^{\alpha}_{\text s, i}=\sigma ^{\alpha}v_i + \alpha _{\text R} \epsilon _{z\alpha i}$ represents the spin-current vertex.

The expectation value of the spin-current density in non-equilibrium states is given by
\begin{align}
\average{\hat j^{\alpha}_{\text s, i}(\bm q)}_{\text{ne},\omega} = \int ^{\infty}_{-\infty} \frac{d\epsilon}{2\pi i} \sum_{\bm k} \text{tr} \left[
j^{\alpha}_{\text s, i} G^< _{\bm k_+,\bm k_-} (\epsilon _+,\epsilon _-)
\right]
,
\end{align}
where $\epsilon _{\pm}=\epsilon \pm \frac{\omega}{2}$.
The trace is taken for the spin space.
Moreover, $G^< _{\bm k_+,\bm k_-} (\epsilon _+,\epsilon _-)$ is the lesser component of the non-equilibrium path-ordered Green function defined by
\begin{align}
G_{\bm k,\bm k'}(t,t') = -i \average{T_C c_{\bm k}(t)c^{\dagger}_{\bm k'}(t')}_{H_{\text T}},
\end{align}
where $T_C$ is a path-ordering operator and $\average{\cdots}_{H_{\text T}}$ represents the expectation value in the non-equilibrium state. 
We expand the lesser Green function for external perturbation $H_{\text R}'$, and calculate the linear response to lattice velocity field $\bm u_{\bm q,\omega}$ up to the first order in $\bm q$.

Considering only the most dominant term (the so-called "Fermi surface term"), the non-equilibrium spin current is given by
\begin{align}
\average{\hat j^{\alpha}_{\text s, i}}^{\text{K/R}}_{\text{ne}} = \frac{i\omega}{2\pi} \sum_{\bm k} \text{tr} \left[
j^{\alpha}_{\text s, i} G^{\text R}_{\bm k}(0) W^{\text{K/R}}_j G^{\text A}_{\bm k}(0)
\right]
u^j_{\bm q, \omega},
\end{align}
where $\average{\cdots}^{\text{K/R}}$ represents the linear response to $H'_{\text K}(H'_{\text R})$.
The results are given by\cite{vcs}
\begin{gather}
\average{\hat j^{\alpha}_{\text s, i}}_{\text{ne}}^{\text K} = 
j_{\text s0} \frac{\tilde \alpha _{\text R}}{4}
\bigl(
4\tilde{c}_2 \eta ^z_{\alpha ijl} + \tilde{a}_4 \eta ^{\text{t}}_{\alpha ijl}
\bigr) iq_l u^j_{\bm q, \omega},
\\
\average{\hat j^{\alpha}_{\text s, i}}_{\text{ne}}^{\text R} = -j_{\text s0} \frac{\tilde \alpha _{\text R}}{4} a_3
 \bigl( 2\epsilon _{\alpha l} \delta ^{\text{t}}_{ij} + 3\epsilon _{\alpha i} \delta ^z_{jl} 
+ \Delta ^{xy}_{\alpha l} \lambda ^{xy}_{ij} 
\bigr) iq_l u^j_{\bm q,\omega},
\label{result}
\end{gather}
where $j_{\text s0}=v_{\text F} \mu \nu _0 \tau$ is a material constant and $\tilde \alpha _{\text R}=\alpha _{\text R}/v_{\text F}$ is the dimensionless Rashba parameter.
In the aforementioned expression, we defined $\delta ^{\text{t}}_{ij}=\delta _{ij}^x+\delta _{ij}^{y}$, $\delta ^i_{jl}=\delta _{ij}+\delta _{il}$, 
$\eta ^{z/\text t}_{\alpha ijl}=\epsilon _{\alpha l}\delta ^{z/\text t}_{ij}+\epsilon _{\alpha j}\delta ^{z/\text t}_{il}+\epsilon _{\alpha i}\delta ^{z/\text t}_{jl}$,
$\Delta ^{xy}_{\alpha l}=\delta _{\alpha x}\delta _{ly}-\delta _{\alpha y}\delta _{lx}$, and $\lambda ^{xy}_{ij}=\delta _{i x}\delta _{jy}+\delta _{iy}\delta _{jx}$.
$a_n=\nu _0^{-1}\sum_{s,\bm k}(s\tilde k_{\text t})^{n-1}\delta (\mu -E_{(s)})$ and $c_n=\nu _0^{-1}\sum_{s,\bm k} \tilde k_z^2(s\tilde k_{\text t})^{n-1}\delta (\mu -E_{(s)})$ were dimensionless parameters (see the Supplementary Materials), with $\tilde k=k/k_{\text F}$ as the dimensionless wavenumber.
For the calculation, we also defined $\tilde a_n=-\tilde \alpha _{\text R}a_n$ and $\tilde c_n = -\tilde \alpha _{\text R} c_n$ for even $n$.
We confirmed that the charge current and spin accumulation did not arise within this calculation; i.e, the spin current was generated directly by the lattice distortion dynamics.
Note that the AC spin current was generated from the present mechanism.

\begin{figure}[tbp]
\centering
\includegraphics[width=60mm]{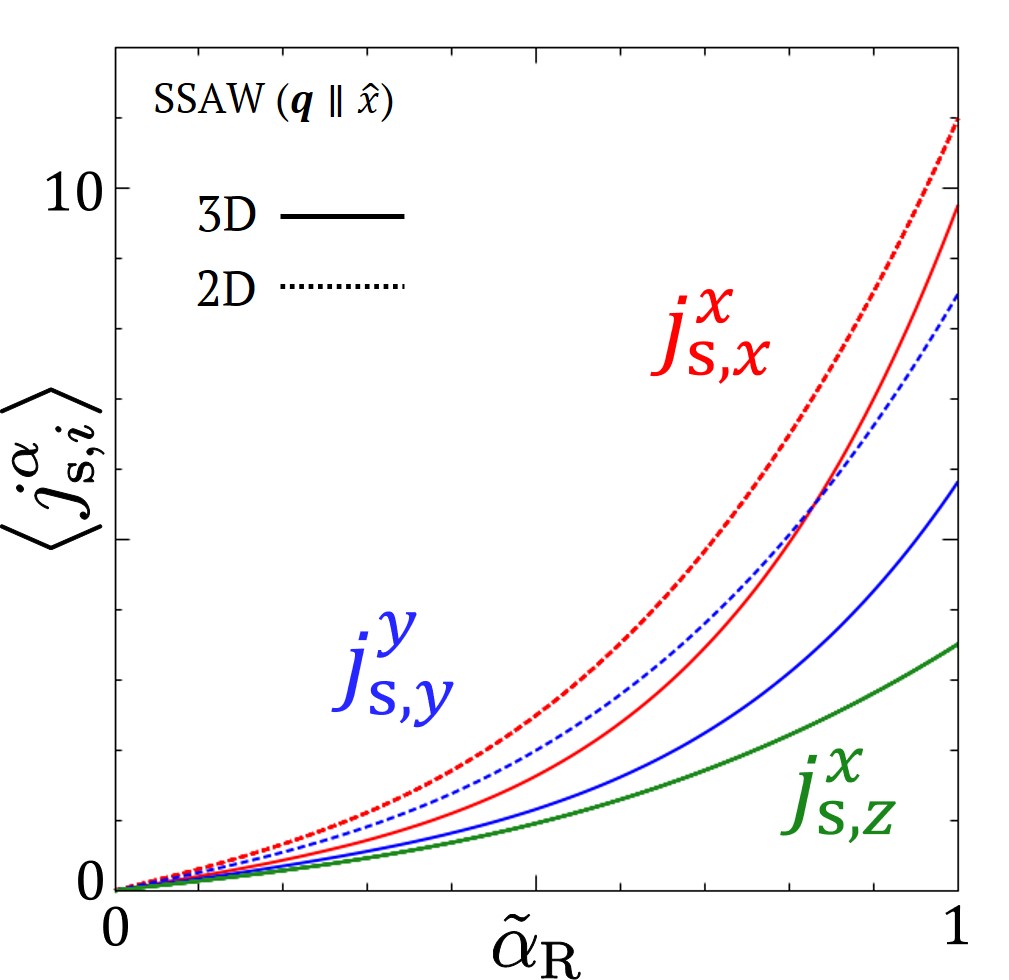}
\caption{
Rashba parameter $\tilde \alpha _{\text R}$ dependence of the spin current generated by the SSAW, which is normalized by $j_{\text s0}qu_0$ or $j_{\text s0}\kappa u_0$.
The red (blue) line represents the helicity current polarized in the $x$ direction ($y$ direction), and the green line represents the perpendicular spin current polarized in the $x$ direction.
The solid (dashed) line represents the spin current generated in 3D (2D) systems.
}
\label{graph}
\end{figure}

\begin{table*}[tbp]
\caption{
Symmetry of the spin current generated by the applied SSAW via each mechanism.
Clearry, the spin current generated via the present mechanism is not generated by conventional mechanisms.
The last line describes SHE case due to the Rashba SOI for reference. 
}
\begin{ruledtabular}
\begin{tabular}{ccccccc}
\multirow{2}{*}{Ref.} & \multirow{2}{*}{Input} & \multirow{2}{*}{Interaction} & Helicity current &\multicolumn{3}{c}{Spin current}
\\ 
& &  & $j^x_x$, $j^y_y$          &      $j^x_z$&$j^y_z$&$j^z_x$ 
\\
\hline \hline
Present & \multirow{2}{*}{SSAW} & Rashba SOI & \Checkmark   & \Checkmark & -- & --
\rule[0mm]{0mm}{3.5mm}
 \\
\cite{funato2018}  &  & Extrinsic SOI & -- & \Checkmark & -- & \Checkmark 
\rule[0mm]{0mm}{3.5mm}
\\ \hline
\cite{matsuo2013} & Rayleigh SAW & SVC  & -- & --  & \Checkmark & --
\rule[0mm]{0mm}{3.5mm}
 \\ 
\cite{sinova2004}  &  Charge current & Rashba SOI & --  & -- & --  & \Checkmark \rule[0mm]{0mm}{3.5mm}
\end{tabular}
\end{ruledtabular}
\label{list}
\end{table*}

We consider a spin current induced via the present mechanism when an SSAW is applied.
The velocity field is given by $\bm u(\bm r, t) = \hat y u_0 e^{-\kappa z} e^{i(qx-\omega t)}$, where $\hat y$ is the unit vector in the $y$ direction, $\kappa$ is the damping constant of the SAW in the $z$ direction, and $u_0$ is the amplitude.
Then, the spin current generated by the SSAW is given by
\begin{gather}
\average{\hat j^{\alpha}_{\text s, i}}^{\text{sw}}_{\text{ne}}
= j_{\text s0} \frac{\tilde \alpha _{\text R}}{8}  \left[ -(2\tilde a_4 +3a_3)\Delta ^{xy}_{\alpha i} + a_3 \lambda _{\alpha i}^{xy}\right] \nabla _xu^y 
\nonumber \\
+j_{\text{s}0} \tilde \alpha _{\text R}\tilde c_2 \delta _{\alpha x}\delta _{iz} \nabla _zu^y.
\end{gather}
The first line is an in-plane helicity current whose spin and flow directions are parallel, 
and the second line is a perpendicular spin current whose spin and flow directions are orthogonal, as illustrated in Fig.~\ref{wave}.
Inparticular, the in-plane helicity current cannot be produced by  conventional methods.
The SVC mechanism produces a spin current whose spin direction is parallel to the vorticity of the lattice,
and the SHE produces a spin current whose spin and flow directions are orthogonal, as listed in Table~\ref{list}.
The reason for such an unconventional spin current is that the present mechanisms produces the spin current directly from the lattice distortion dynamics via the Rashba SOI without a charge current or spin accumulation.
However, the spin current flowing to the $z$ direction is along the SAW decay, which is generated only in the case of the 3D systems.
This current can also be induced via the extrinsic SHE, but cannot be understood by the Rashba SHE because the Rashba SHE produces only a $z$ polarized spin current\cite{sinova2004}.

Let us estimate the generated spin current in specific materials:
For a Cu/Bi$_2$O$_3$ bilayer\cite{karube2016}, the Rashba parameter is $\tilde \alpha _{\text R}= 0.1$.
With material parameters of bulk Cu (Fermi energy $\mu \sim 7.03$[eV], Fermi velocity $v_{\text F} \sim 1.57\times 10^6$[m/s], density of states $N_0(\mu ) = 1.81\times 10^{28}$[m$^{-3}$eV$^{-1}$], and relaxation time $\tau \sim 10^{-15}$[s]), we estimate $j_{\text s0}qu_0 \sim 3.0 \times 10^{27}$[m$^{-2}$s$^{-1}$] for lattice displacement $\delta R\sim 1$\AA, frequency $f\sim 3.8$[GHz], and phase velocity $v_{\text a} \sim 3.8$[km/s].
The spin currents have magnitudes that is large enough to be observable. 
However, they are expected to be smaller than $j_{\text s0}qu_0$ obtained here because of the spin current flowing throughout the bulk.
Regarding BiTeI as a 3D Rashba system with $\tilde \alpha _{\text R}\sim 1$\cite{demko2012,shibata2018}, the normalized parameter for the helicity current is estimated as $j_{\text s0}qu_0 \sim 6\times 10^{16}$-$10^{19}$[m$^{-2}$s$^{-1}$] with $\mu =0.2$[eV], $v_{\text F}\sim 10^{5}$[ms$^{-1}$], $N_0(\mu )\sim 3\times 10^{26}$[m$^{-3}$eV$^{-1}$], and $\tau \sim 10^{-15}$-$10^{-18}$[s].
The normalized parameter for the orthogonal spin current can be estimated as $j_{\text s0}\kappa u_0\sim j_{\text s0}qu_0$ because damping constant $\kappa$ is approximately equal to the wavenumber $\kappa \sim q$.
As a detection of AC spin currents, spin wave resonances have been observed by injecting spin currents into ferromagnetic metals\cite{kobayashi2017,tateno2020}.
Moreover, AC spin currents have been detected by the rectification effect due to magnetostriction\cite{kawada2021}.
The spin current produced by the present mechanism may be observed by using these methods.

In summary, we studied the spin current generation due to dynamical lattice distortion in 2D and 3D Rashba electron systems using nonequilibrium quantum field theory.
We found a process in which spin currents are generated directly by the lattice distortion dynamics. Because of the interplay between the Rashba SOI and the dynamical lattice distortion, unlike conventional spin transport, the in-plane helicity currents are generated without an accompanying charge current or spin accumulation. 
We also found that the in-plane helicity currents do not flow parallel to the vorticity of the lattice distortions.
The obtained spin currents were experimentally detectable in strong Rashba systems and are expected to be able to be observed by spin-wave resonance. Overall, our theory revealed alternative spintronic functionalities of Rashba systems.

We would like to thank M. Hayashi, T. Kawada, Y. Nozaki, K. Yamanoi, and T. Horaguchi for enlightening discussions.
We also thank H. Kohno, A. Yamakage, Y. Imai, J. J. Nakane. Y. Yamazaki, Y. Ogawa, Y. Ozu, and Y. Hayakawa for daily discussions.
This work was partially supported by JST CREST Grant Number JPMJCR19J4, Japan.
TF is supported by Grant-in Aid for JSPS Fellows Grant Number 19J15369, and by a Program for Leading Graduate Schools "Integrative Graduate Education and Research in Green Natural Sciences". MM is supported by Grant-in-Aid for Scientific Research B (20H01863).

\end{document}